\documentclass[final]{hubsarxiv}
\graphicspath{{figures/}}
\usepackage{tabularx}
\usepackage{threeparttable}
\usepackage{listings}
\lstdefinelanguage{Turtle}{
  morekeywords={a, @prefix, @base},
  sensitive=true,
  morecomment=[l]{\#},
  morestring=[b]",
}
\title{BuildGraph: A Synthetic Multi-Archetype Building Knowledge Graph Dataset}
\shorttitle{BuildGraph: a synthetic multi-archetype building knowledge graph dataset}
\shortauthors{Wooyoung Jung}
\author{Wooyoung Jung\affmark{a,}\corrmark\orcid{0000-0003-2746-4983}}
\affil{\affmark{a}Department of Civil and Architectural Engineering and Mechanics, University of Arizona, Tucson, AZ, USA}
\corresponding{wooyoung@arizona.edu}
\paperno{2026-005}
\paperdate{September 2026}
\versionnote{v1 (2026-09-02): first release. v2 (2026-11-05): re-set in the HUBS Lab paper layout; content unchanged.}
\license{CC BY-NC-ND 4.0}
\abstract{Semantic querying of building knowledge graphs (KGs) underpins the integration of artificial
intelligence into building operations, from natural-language access to cross-building analytics, 
but such KGs are rarely public owing to proprietary, security, and cost barriers. BuildGraph is 
a synthetic building KG dataset of 120 buildings in the Brick Schema ontology, grounded in U.S.\ 
Department of Energy prototype models and sensor-placement patterns from real buildings. 
It spans eight commercial building types across three ASHRAE energy-code vintages, with five 
realizations per archetype varying URI naming, sensor-attachment predicates, and topology. 
A 75-query SPARQL benchmark confirms structural completeness: BuildGraph reaches 91.5\% Query 
Answerability Rate versus 35.8\% for 59 real-world Brick files. Independently, it reproduces real buildings'
sensor-type proportions on their shared vocabulary (cosine 0.86--0.94), evidence of realistic
instrumentation where measurable. A downstream text-to-SPARQL experiment with Gemma~4 (26B)
reaches 27.3\% Row-Matching~F1 (+17.5~pp over zero-shot) on 12 held-out buildings. BuildGraph gives
facility managers and digital-twin developers a testbed for portable analytics and natural-language
interfaces, and its dataset, generator, and benchmark are openly available at
\url{https://github.com/humanbuildingsynergy/BuildGraph}.}

\begin{document}
\maketitle

\section*{Practical Applications}

BuildGraph gives building practitioners a free, ready-made set of 120 realistic building knowledge
graphs and a 75-question query benchmark that would otherwise require proprietary, hard-to-obtain
building data. Facility managers and controls engineers can test fault-detection and
energy-analytics queries across many building types and vintages before applying them to their own
sites; software developers can benchmark tools that translate plain-English questions into
building-database queries; and digital-twin developers can prototype and validate semantic
applications without first assembling a real building's knowledge graph. Because the generator is
released, practitioners can also produce additional buildings tailored to their own portfolio.

\section{Introduction} 

Building knowledge graphs (KGs) are semantic models that represent the physical components,
relationships, and instrumentation of a building as a machine-readable graph. For example,
a building KG encodes which air handling units (AHUs) serve which heating, ventilation, and 
air conditioning (HVAC) zones, which sensors are attached to which equipment, and how chilled 
water and hot water flow through the plant -- expressed in a formal ontological vocabulary 
such as the Brick Schema~\citep{balaji2018brick,fierro2022brick}. This structured representation underpins
the integration of artificial intelligence into building operations: it enables the semantic
querying, inference, and reasoning behind applications that range from natural-language
interfaces---which let practitioners retrieve building information without SPARQL expertise---to
portable analytics that transfer across buildings. As the building industry increasingly embraces semantic models, building 
KGs are becoming a critical data infrastructure, and their public availability is essential for 
research progress.

However, the research community faces a severe data scarcity problem. Real-world building KGs
are rarely publicly available due to proprietary restrictions, security concerns, and lack of
instrumented buildings, limiting supply to a small number of anonymized KG models released by
national laboratories and universities. This scarcity is particularly acute for text-to-SPARQL
research -- systems that translate natural language questions into executable SPARQL Protocol 
and RDF Query Language (SPARQL) queries, emerging as a critical interface for facility managers 
and energy engineers to access building knowledge without SPARQL expertise -- because such 
systems require diverse, structurally complete KGs as few-shot examples, training data, and 
evaluation benchmarks. The only existing benchmark, BuildingQA~\citep{buildingqa}, covers four
buildings annotated in a mixture of Brick, ASHRAE~223P, and mixed-schema formats, providing 
insufficient scale and structural diversity for systematic evaluation or cross-building 
generalization studies.

This data scarcity problem is likely to persist for the foreseeable future due to a steep
learning curve for practitioners to learn, create, and maintain building
KGs~\citep{bhattacharya2015automated,koh2018plaster}, and the fact that many buildings will
not be fully instrumented or semantically modeled for years to come. Even if more real 
building KGs are released, they will likely be limited in number, biased toward certain 
building types or instrumentation patterns, and encumbered by proprietary restrictions that 
limit their utility for research and evaluation.

Synthetic generation of building KGs is a promising strategy to address this data scarcity
problem and provide a diverse, scalable, and publicly accessible dataset for
research~\citep{nikolenko2021synthetic}. Structurally realistic synthetic KGs -- matching the
equipment types, sensor placement patterns, topological relationships, and HVAC system
configurations found in real buildings -- can substitute for real KGs as few-shot examples
and evaluation benchmarks without privacy risk or access barriers. The key design challenge
is ensuring that the synthesis is empirically grounded rather than arbitrary: a randomly
generated graph can satisfy syntactic constraints while being physically implausible, producing
a dataset that teaches a model the wrong distribution.

This data paper introduces \textbf{BuildGraph} and its associated dataset: a collection of 
120 synthetic building KGs in the Brick Schema ontology, grounded in two independent empirical 
sources. First, building configurations are derived from the U.S.\ Department of Energy (DOE) 
Commercial Prototype Building Models~\citep{doe2024prototype}, published specifications that 
define the HVAC system type, equipment counts, terminal unit configuration, and zone structure 
for eight representative commercial building types spanning three energy code vintages. These 
models have been validated against national energy survey data and are the standard reference 
for U.S.\ commercial building simulation research. Second, sensor placement probabilities are 
calibrated against real buildings from the Mortar corpus~\citep{fierro2019mortar,mortar_resources}, 
ensuring that the frequency and type of sensors attached to each equipment class reflects observed 
practice across real instrumented buildings.

Dataset quality is assessed along two dimensions. First, a 75-query SPARQL benchmark measures
the Query Answerability Rate (QAR) -- the fraction of applicable queries returning at least one
result row -- across five query categories and four building complexity tiers. BuildGraph
achieves an overall QAR of 91.5\%, compared to 35.8\% for a corpus of 59 publicly available
real-world Brick building files -- indicating that the generated graphs are substantially more
structurally complete with respect to this benchmark. Second, a downstream text-to-SPARQL 
experiment directly evaluates the dataset's utility for cross-building generalization.

The primary contributions of this paper are:

\begin{enumerate}[leftmargin=0.75in]

  \item A publicly accessible dataset of \textbf{120 synthetic building KGs} in the Brick Schema
  ontology, spanning eight commercial building types (small office, medium office, large office,
  primary school, hospital, hotel, strip mall, and outpatient healthcare), three ASHRAE energy
  code vintages, and various HVAC configurations; archetype coverage validated against the 
  Commercial Buildings Energy Consumption Survey (CBECS) 2018 data -- a national sample survey
  that details the stock of U.S. commercial buildings. This dataset provides a critical resource for 
  text-to-SPARQL research and building ontology benchmarking.

  \item A \textbf{75-query SPARQL benchmark} for measuring structural quality of building KGs,
  with scope-detection logic that automatically determines which queries are applicable to each
  graph and an evaluation framework aligned with building complexity.

  \item A documented, open-source \textbf{methodology for reproducible synthetic building KG
  generation}, grounded in published prototype models and empirical sensor placement data,
  enabling the research community to extend the dataset to additional building types, ontology
  modes, or seeds.

  \item \textbf{Downstream validation} via a text-to-SPARQL evaluation using Gemma~4
  (26B)~\citep{gemma4} on 12 held-out buildings, demonstrating that few-shot prompting with
  cross-building BuildGraph text-to-SPARQL examples achieves 27.3\% Row-Matching~F1
  (the harmonic mean of precision and recall over the \textit{set of result rows} a query returns---
  measuring whether the query returns the correct answer independent of SPARQL surface form, and
  defined in Section~\ref{sec:downstream}; +17.5~pp over zero-shot), providing a baseline
  performance using BuildGraph examples.

\end{enumerate}

\section{Background}

\subsection{From Sensor Data to Queryable Knowledge Graphs}

\begin{figure}[htbp]
  \centering
  \includegraphics[width=0.60\linewidth]{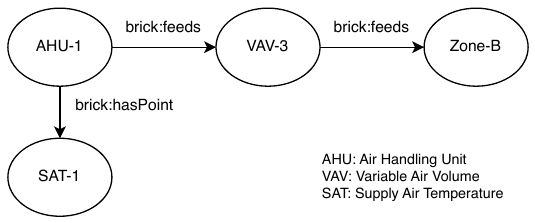}
  \caption{Example building knowledge graph fragment. Nodes (ovals) are building
           entities; labeled arrows are typed semantic relationships. Following the two
           \texttt{brick:feeds} arrows in sequence -- from AHU-1 through VAV-3 to
           Zone-B -- is an example of a multi-hop query.}
  \label{fig:rdf_example}
\end{figure}

Integrating heterogeneous building sensor and equipment data across vendors and sites
requires a common representation in which entities, their types, and the relationships
between them are explicitly encoded. The Resource Description Framework
(RDF)~\citep{w3c_rdf} provides this representation through a simple but expressive model:
every fact is a \textit{triple} of the form \textit{subject--predicate--object}. For
example, the triple (\texttt{AHU-1}, \texttt{brick:feeds}, \texttt{VAV-3}) states that
AHU-1 supplies conditioned air to VAV-3, while (\texttt{AHU-1}, \texttt{brick:hasPoint},
\texttt{SAT-1}) attaches a supply air temperature sensor to that unit. Assembling many such
triples produces a directed, labeled graph -- a \textit{knowledge graph} (KG) -- in which
nodes represent building entities and edges represent typed semantic relationships
(Figure~\ref{fig:rdf_example}).

Shared meaning requires a common vocabulary: the Brick Schema~\citep{balaji2018brick}
serves as this shared ontology, defining entity type hierarchies (e.g., \texttt{brick:AHU}
as a subclass of \texttt{brick:HVAC\_Equipment}) and the formal semantics of each
relationship, while each building's RDF file encodes its own instance data.
SPARQL~\citep{sparql11} is the W3C standard query language for RDF, enabling queries to
pattern-match against building instance triples using Brick class and property definitions
to navigate relationships. A key capability is the property path operator (\texttt{brick:feeds+}), 
which matches one or more consecutive \texttt{brick:feeds} edges, enabling a single query 
to trace the full supply chain from chiller to zone. Such multi-hop traversals 
require a \textit{structurally complete} A-box: every expected entity and relationship must be present.

\subsection{Schema Conformance vs.\ Query Answerability}

Structural completeness cannot be measured by schema validation alone. The Shapes Constraint
Language (SHACL)~\citep{shacl2017} is the W3C standard for validating whether an RDF
graph conforms to the property definitions and cardinality constraints declared in the T-box:
for example, that every \texttt{brick:TemperatureSensor} has the required
\texttt{brick:isPointOf} relationships to entities of type \texttt{brick:Equipment}
(\texttt{brick:isPointOf} is the sensor-to-equipment direction; its inverse,
\texttt{brick:hasPoint}, attaches sensors from the equipment side).
Brick~1.3~\citep{fierro2022brick} introduced SHACL shapes that enable automated conformance
checking for any Brick model. Passing SHACL is a necessary condition for a well-formed
KG -- but it is not sufficient.

\begin{figure}[htbp]
  \centering
  \includegraphics[width=0.95\linewidth]{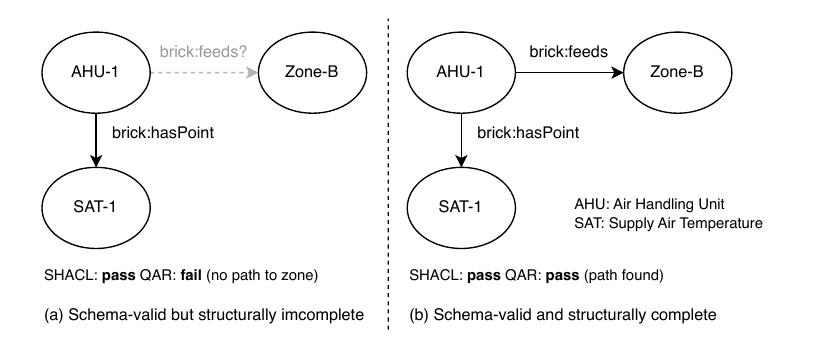}
  \caption{SHACL conformance verifies that present relationships are schema-valid;
           QAR measures whether the graph contains the relationships that queries
           require. Both graphs above satisfy SHACL, but only (b) answers a query
           asking which zones an AHU serves.}
  \label{fig:shacl_vs_qar}
\end{figure}

A KG can satisfy every SHACL constraint and still fail to answer queries that practitioners
care about. A building model might correctly type all equipment nodes and satisfy all
cardinality rules, yet omit the \texttt{brick:feeds} chain connecting AHUs to zones, the
\texttt{brick:hasPart} relationships linking sub-components, or the zone-level sensor
attachments needed for occupancy and thermal comfort queries. These omissions are invisible
to SHACL because SHACL checks \textit{conformance} -- whether what is present is
schema-valid -- not \textit{completeness} -- whether everything a query requires is there.
The appropriate measure is therefore query-centric: the \textit{Query Answerability Rate}
(QAR), defined as the fraction of applicable, semantically meaningful SPARQL queries that
return at least one result row when executed against a building KG. A high QAR confirms that
the A-box is structurally rich enough to support the full range of queries a downstream
application would issue; a low QAR reveals coverage gaps -- analogous to the
\textit{query-based completeness} dimension in linked data quality literature~\citep{zaveri2016quality}
-- that would silently produce empty results in practice
(Figure~\ref{fig:shacl_vs_qar}). In operational terms, structural completeness
translates directly into computational reliability: a query issued by a live
building analytics application against a structurally complete graph returns the
expected result rows rather than silently failing, reducing the query-execution
failures that arise when required equipment or sensor relationships are absent.

QAR measures \textit{query-based completeness}: whether the relationships a workload 
of queries requires are present in the graph. It is therefore a necessary condition for 
a dataset that must support semantic querying, but it is not a sufficient condition for 
dataset quality along three axes. First, QAR does not measure the \textit{semantic correctness} 
of individual triples: a query returning a non-empty result confirms that a matching 
pattern exists, not that the underlying assertion is physically accurate. Second, QAR
does not measure \textit{realism}: a graph can be structurally complete yet instrument 
its equipment in ways that do not reflect deployed practice. Third, QAR is defined \textit{relative 
to a chosen query workload}; a graph modeled for a narrower purpose may score low despite 
being valid for that purpose, and the metric inherits whatever coverage the benchmark 
queries provide. QAR therefore serves as the central metric for the property it is designed to measure,
structural completeness, and is complemented---not replaced---by independent assessments of
sensor-placement realism, archetype representativeness, and downstream utility that address
the dimensions beyond its scope. 

\section{Related Work}

\subsection{Building Metadata Standards and the Brick Schema}

Early standardization efforts addressed communication interoperability (BACnet~\citep{bacnet1995})
and tag-based operational semantics (Project Haystack~\citep{haystack2014}), but lacked typed,
named relationships between entities, making SPARQL-based querying and ontological inference
impossible~\citep{bhattacharya2015schemas}. Project Haystack 4 attempted to address this by
layering Web Ontology Language (OWL) structure on the tag model, acknowledging that the folksonomy 
approach was insufficient for programmatic reasoning~\citep{pritoni2021metadata} -- yet 
retrofitting formal semantics onto a flat tagging system proved architecturally incomplete.
\citet{balaji2018brick} resolved this gap from first principles, designing a formal OWL
ontology whose class hierarchy and relationship vocabulary were derived from empirical analysis
of real building analytics requirements. Brick encodes the full building system as a typed RDF
graph structured around three perspectives: \emph{composition} (\texttt{brick:hasPart}),
\emph{topology} (\texttt{brick:feeds}), and \emph{telemetry}
(\texttt{brick:hasPoint}/\texttt{brick:isPointOf}) -- enabling the multi-hop SPARQL traversals
that tag-based schemas could not express.

The value of Brick at scale was demonstrated by the Mortar project~\citep{fierro2019mortar},
which published anonymized Brick models for 45 real buildings from multiple U.S.\ university
campuses and provided a unified SPARQL query interface over building data. Mortar showed that 
a single set of portable analytics applications -- fault detection rules, occupancy-based 
control algorithms, energy benchmarking queries -- could be executed across heterogeneous 
buildings without modification~\citep{balaji2018apenergy}, validating Brick's core design goal.
A subsequent review of 40 building metadata schemas by \citet{pritoni2021metadata} confirmed
that Brick offers the strongest foundation for semantic querying among existing schemas. 
Complementary standards have since emerged -- ASHRAE~223P~\citep{ashrae223p}
for connection-point topology, the Building Topology Ontology (BOT)~\citep{rasmussen2021bot}
for spatial decomposition within the W3C Linked Building Data community, and
RealEstateCore~\citep{realestatecore} for spatial and asset management layers -- but Brick
remains the de facto ontology for building analytics research, and all major public building
KG resources are Brick-based.

Despite its maturity, the supply of publicly available Brick KGs is severely limited.
Table~\ref{tab:public_datasets} summarizes the principal public building KG resources.
The Mortar project~\citep{fierro2019mortar,mortar_resources} is the largest of these resources,
offering publicly downloadable Brick models of 45 real buildings, but as a fixed collection of
deployed real models released without a paired query benchmark. The Open223
project~\citep{open223} provides approximately 19 unique 223P example buildings from national
laboratories, freely available but a small, illustrative set that likewise ships no
query benchmark. The BuildingQA benchmark~\citep{buildingqa} is the only resource providing
paired (natural language, SPARQL) question-answer data, yet covers just 4 buildings. This
scarcity -- driven by proprietary restrictions, security concerns, and the expert effort
required to construct and validate a Brick KG~\citep{bhattacharya2015automated,koh2018plaster}
-- is a recognized bottleneck for training and evaluating text-to-SPARQL systems and is the
primary motivation for this work.

\begin{table}[htbp]
  \centering
  \caption{Public building knowledge graph resources. QA pairs refers to
           (natural language question, SPARQL query) annotation pairs.}
  \label{tab:public_datasets}
  \begin{tabular}{llll}
    \toprule
    \textbf{Resource} & \textbf{Buildings} & \textbf{Ontology} & \textbf{QA Pairs} \\
    \midrule
    Mortar~\citep{fierro2019mortar,mortar_resources}   & 45       & Brick              & None \\
    Open223~\citep{open223}           & $\sim$19 & 223P               & None \\
    BuildingQA~\citep{buildingqa}     & 4        & Brick, 223P, Mixed & Yes \\
    \bottomrule
  \end{tabular}
\end{table}

\subsection{Synthetic Building Data Generation}

Prior work on generating building models falls into two categories distinct from semantic KG
generation. Building information modeling (BIM) tools produce models in the Industry Foundation
Classes (IFC) format~\citep{ifc2013} capturing geometry and spatial hierarchy but not sensor 
attachments or typed semantic relationships; energy simulation tools such as 
EnergyPlus~\citep{crawley2001energyplus} produce thermodynamic outputs rather than RDF graphs.

A parallel line of work infers semantic metadata from \emph{existing} building automation system 
(BAS) deployments via signal processing and active learning~\citep{gao2015metadata}, semi-automated 
metadata normalization~\citep{koh2017metadata}, and neural point name 
classification~\citep{almashor2023pointnames}. Despite strong per-building accuracy, construction 
cost remains high -- hours to days of expert effort per building~\citep{bhattacharya2015automated,koh2018plaster}
-- and all methods require existing building data as input rather than generating
KGs from scratch. Moreover, these learning-based approaches -- including graph-neural models for 
building semantic enrichment~\citep{wang2022gnn} and sequence models for schema matching and metadata
normalization -- share a dependence on diverse, labeled building graphs for training and
evaluation, which today are typically small and hand-compiled (for instance,
\citet{wang2022gnn} train on 224 apartment layouts). Their progress is therefore constrained by
the same data scarcity this work targets: a large, structurally consistent synthetic corpus
offers one route to the scale and variety such methods require~\citep{nikolenko2021synthetic}.

Synthetic data generation is well established in adjacent fields -- from synthetic KGs and question-answering 
corpora to simulated sensor streams -- but building informatics has lacked an equivalent resource.

\subsection{Text-to-SPARQL for Building Systems}

Text-to-SPARQL has been studied extensively for general-purpose KGs, with large-scale
benchmarks including Spider~\citep{yu2018spider}, LC-QuAD~\citep{trivedi2017lcquad,dubey2019lcquad2},
and recent LLM-based approaches achieving high accuracy on open-domain
graphs~\citep{kovriguina2024sparql,lan2021survey}.

Adaptation to building KGs introduces domain-specific challenges absent from general
benchmarks. HVAC topology requires multi-hop \texttt{brick:feeds+} queries that traverse
chained equipment connections; optional equipment classes such as energy recovery ventilators
and cooling towers are present only in some building types; and real deployments exhibit
heterogeneous sensor attachment vocabularies, mixing \texttt{brick:hasPoint} and
\texttt{brick:isPointOf} across buildings and tools. These challenges mean that models
trained on general KG benchmarks do not transfer directly to building KGs, and
domain-specific data at sufficient scale -- as few-shot examples, training material, or
evaluation benchmarks -- is required.

The existing public building-KG resources (Table~\ref{tab:public_datasets}) each leave a different
gap for method development. BuildingQA pairs natural-language questions with SPARQL but covers only
four buildings; Mortar publishes 45 real Brick models, but as a fixed, uncontrolled collection with
no paired query benchmark; and Open223 offers roughly nineteen illustrative example buildings,
likewise without one. What none provides is a \emph{controlled} corpus---a balanced, labeled set of
buildings in which building type, vintage, naming convention, and sensor-attachment predicate are
known and can be varied one factor at a time. BuildGraph is built to fill this gap: it spans a
balanced $8\times3\times5$ factorial of eight building types, three vintages, and five independent
seeds (120 buildings), each released with its full provenance (archetype, vintage, seed) and exact
equipment and sensor inventory, and with naming scheme, attachment predicate, and topology varied
across seeds \emph{by design}---the same heterogeneity real corpora exhibit, but here labeled and
reproducible. It ships with a reusable 75-query QAR benchmark that runs across all 120 buildings,
and, because the generator is released, it is not a fixed corpus but an open, extensible resource:
users can create additional buildings---new archetypes, vintages, sizes, and naming or predicate
conventions---at any scale. To the author's knowledge it is the first large-scale synthetic
building-KG dataset released with an accompanying evaluation protocol, giving research groups an
identical, citable, and extensible substrate on which to develop and reproducibly evaluate
methods---the shared infrastructure that portable building-analytics research currently lacks.

\section{Dataset Description}

\subsection{Overview}

The BuildGraph dataset comprises 120 synthetic building knowledge graphs in the Brick Schema
ontology~\citep{balaji2018brick,fierro2022brick}, serialized in RDF Turtle (TTL) format~\citep{turtle2014}.
The 120 buildings are produced by generating five independent realizations of each of the
24 building archetypes defined by crossing eight commercial building types with three ASHRAE
energy code vintages. Each realization introduces controlled, reproducible variation in
building structure, naming conventions, and sensor placement, producing structurally
distinct instances of the same archetype (Section~\ref{sec:archetypes}).

Each building file is self-contained: all entity declarations, relationship triples, sensor
attachment triples, and building metadata are contained in a single file. No external ontology
files are required; standard Brick prefix declarations are embedded in every file.
Dataset scale ranges from 81 triples (a minimally instrumented pre-1980 small office)
to 12,790 triples (a dense 2013-vintage hotel with full central plant
topology), with an overall mean of 2,107 triples and 451 typed entities per building.
Table~\ref{tab:summary} summarizes the dataset at a glance.

\begin{table}[htbp]
\centering
\caption{BuildGraph dataset summary.}
\label{tab:summary}
\begin{tabular}{ll}
\toprule
Property & Value \\
\midrule
Buildings (files)      & 120 (8 types $\times$ 3 vintages $\times$ 5 seeds) \\
Ontology               & Brick Schema 1.3 \\
Serialization          & RDF Turtle (\texttt{.ttl}), one self-contained file per building \\
Triples per building   & 81 (min) / 2{,}107 (mean) / 12{,}790 (max) \\
Typed entities/building & 20 (min) / 451 (mean) / 2{,}331 (max) \\
Distinct Brick classes & 83 \\
SPARQL benchmark       & 75 queries, 5 categories, 14 scope checks \\
License                & CC BY-NC 4.0 \\
Repository / DOI        & \url{https://doi.org/10.5281/zenodo.20015123} \\
\bottomrule
\end{tabular}
\end{table}

\subsection{Building Archetypes and Vintages}
\label{sec:archetypes}

The 24 archetypes are defined by crossing eight commercial building types with three ASHRAE
energy code vintages. Building types and their HVAC configurations are summarized in
Table~\ref{tab:archetypes} and Table~\ref{tab:vintages}. The combined effect of building 
type and vintage on dataset scale is summarized in Table~\ref{tab:dataset_stats}; values are means 
over the five realizations per archetype.

\begin{table}[htbp]
  \centering
  \begin{threeparttable}
    \caption{BuildGraph archetype coverage.}
    \label{tab:archetypes}
    \begin{tabular}{llll}
      \toprule
      \textbf{Building Type} & \textbf{HVAC System} & \textbf{Terminal Units} &
      \textbf{Central Plant} \\
      \midrule
      Small office        & PSZ-AC (packaged RTU) & CAV          & No  \\
      Medium office       & PVAV w/ reheat        & VAV mix      & Yes \\
      Large office        & VAV w/ reheat         & VAV mix      & Yes \\
      Primary school      & PSZ-AC (packaged RTU) & CAV          & No  \\
      Hospital            & VAV + DOAS            & VAV reheat   & Yes \\
      Hotel               & FCU / DOAS+FCU (2013) & FCU          & Yes \\
      Strip mall          & PSZ-AC (packaged RTU) & CAV          & No  \\
      Outpatient health.\ & FCU / DOAS+FCU        & FCU          & Yes \\
      \bottomrule
    \end{tabular}
    \begin{tablenotes}[flushleft]
      \footnotesize
      \item[] PSZ-AC = Packaged Single Zone, Air-Cooled; PVAV = Packaged Variable Air Volume;
              VAV = Variable Air Volume; CAV = Constant Air Volume;
              DOAS = Dedicated Outdoor Air System; FCU = Fan Coil Unit; RTU = Rooftop Unit.
    \end{tablenotes}
  \end{threeparttable}
\end{table}

\begin{table}[htbp]
  \centering
  \caption{Vintage effects on sensor density, metering, and system configuration.}
  \label{tab:vintages}
  \begin{tabular}{lccc}
    \toprule
    \textbf{Dimension} & \textbf{pre-1980} & \textbf{90.1-2004} & \textbf{90.1-2013} \\
    \midrule
    Sensor density scale & 0.40$\times$ (sparse) & 1.00$\times$ (standard) &
                           1.40$\times$ (dense) \\
    Electric metering    & No  & Yes & Yes \\
    Heat recovery (DOAS) & No  & No  & Yes (ERV on all DOAS AHUs) \\
    \bottomrule
  \end{tabular}
\end{table}

\begin{table}[htbp]
  \centering
  \begin{threeparttable}
    \caption{BuildGraph dataset scale by building type and vintage (means over 5 realizations).}
    \label{tab:dataset_stats}
    \begin{tabular}{lrrrrrr}
      \toprule
      & \multicolumn{2}{c}{\textbf{pre-1980}} &
        \multicolumn{2}{c}{\textbf{90.1-2004}} &
        \multicolumn{2}{c}{\textbf{90.1-2013}} \\
      \cmidrule(lr){2-3} \cmidrule(lr){4-5} \cmidrule(lr){6-7}
      \textbf{Building Type} &
        \textbf{T} & \textbf{E} & \textbf{T} & \textbf{E} & \textbf{T} & \textbf{E} \\
      \midrule
      Small office        &   181 &  42 &   395 &   86 &   479 &  105 \\
      Medium office       &   421 & 102 & 1,192 &  270 & 1,440 &  322 \\
      Large office        &   923 & 219 & 2,686 &  600 & 3,203 &  708 \\
      Primary school      &   511 & 119 & 1,546 &  340 & 1,848 &  404 \\
      Hospital            & 1,602 & 367 & 4,861 & 1,015 & 6,566 & 1,372 \\
      Hotel               & 2,277 & 461 & 5,280 & 1,097 & 7,540 & 1,547 \\
      Strip mall          &   289 &  69 &   817 &  183 & 1,324 &  291 \\
      Outpatient health.\ &   593 & 131 & 1,705 &  359 & 2,880 &  608 \\
      \bottomrule
    \end{tabular}
    \begin{tablenotes}[flushleft]
      \footnotesize
      \item[] T = RDF triples; E = typed \texttt{brick:} entity instances.
    \end{tablenotes}
  \end{threeparttable}
\end{table}

\textbf{Variation across realizations.}
Five structurally distinct yet consistent instances are generated per archetype across three
controlled dimensions.

\textbf{(1) Building structure.}
Zone count, floor count, AHU count, zone type assignments, zone areas, climate zone, and
equipment configuration details (economizer and exhaust dampers, number of chillers and
boilers) are drawn within the ranges specified by the DOE prototype model for that archetype;
sensor inclusion per equipment instance is sampled from empirical base probabilities derived
from the Mortar corpus~\citep{fierro2019mortar} and scaled by the vintage density factor.

\textbf{(2) Entity naming.}
Each realization independently draws a Uniform Resource Identifier (URI) naming convention 
from four styles weighted to
reflect patterns observed in deployed building automation systems: \textit{numeric}
(e.g., \texttt{bldg:AHU01}) predominates because vendor-assigned numeric tags are the
default in most BAS platforms; \textit{hyphenated} (\texttt{bldg:AHU-01-North}) and
\textit{descriptive} (\texttt{bldg:MainAirHandlerUnit}) appear in site-specific and manually
curated models; and \textit{legacy underscore} (\texttt{bldg:AHU\_01\_Supply}) is retained
for compatibility with older deployments, preventing a model from exploiting systematic
identifier patterns.

\textbf{(3) Sensor attachment predicate.}
Each realization independently selects whether sensor-to-equipment attachment uses
\texttt{brick:hasPoint} exclusively (50\%), \texttt{brick:isPointOf} exclusively (25\%),
or both simultaneously (25\%), reflecting the predicate heterogeneity observed across
real Brick deployments.

\subsection{Knowledge Graph Content}

\subsubsection{Entity Classes}

The dataset uses 83 distinct Brick classes. Table~\ref{tab:entity_classes} organizes
the equipment entity classes by functional category.

\begin{table}[tbp]
  \centering
  \caption{Equipment entity classes in BuildGraph by functional category.
           ``Scope'' indicates which buildings include instances of that category.}
  \label{tab:entity_classes}
  \begin{tabular}{p{3.5cm} p{6.5cm} p{3.0cm}}
    \toprule
    \textbf{Category} & \textbf{Brick classes} & \textbf{Scope} \\
    \midrule
    Building hierarchy
      & \texttt{Building}, \texttt{Floor}, \texttt{HVAC\_Zone}
      & All \\
    Air-handling units
      & \texttt{AHU}, \texttt{Supply\_Fan}, \texttt{Return\_Fan},
        \texttt{Heating\_Coil}, \texttt{Cooling\_Coil},
        \texttt{Variable\_Frequency\_Drive},
        \texttt{Economizer\_Damper}, \texttt{Exhaust\_Damper},
        \texttt{Return\_Damper}
      & All \\
    Heat recovery
      & \texttt{Heat\_Exchanger} (ERV)
      & 2013 DOAS only \\
    VAV terminals
      & \texttt{VAV},
        \texttt{Variable\_Air\_Volume\_} \texttt{Box\_With\_Reheat}
      & VAV, PVAV, and CAV systems \\
    FCU terminals
      & \texttt{Fan\_Coil\_Unit}
      & FCU systems \\
    Central plant
      & \texttt{Chiller}, \texttt{Cooling\_Tower}, \texttt{Boiler},
        \texttt{Chilled\_Water\_Pump}, \texttt{Hot\_Water\_Pump},
        \texttt{Condenser\_Water\_Pump}
      & Central-plant buildings \\
    Metering
      & \texttt{Electrical\_Meter}
      & 2004 and 2013 \\
    \bottomrule
  \end{tabular}
\end{table}

\subsubsection{Sensor and Control Points}

Sensor and control points span four domains, sampled from probabilities calibrated against the
Mortar dataset. Zone-level points include air temperature, CO$_2$, occupancy, and air quality
sensors and setpoints. AHU-level points cover duct air temperatures (supply, return, discharge,
mixed, outside air), flow, humidity, static pressure, filter differential pressure, and AHU
control commands. VAV and FCU terminals carry discharge air temperature, airflow, airflow
setpoint, and damper and valve commands. Central-plant equipment carries supply and return
temperature, flow, and differential pressure for both hydronic loops, plus power and gas
metering sensors.

\subsubsection{Building and Zone Metadata}

Each building node carries three metadata triples: \texttt{meta:buildingType} records the
archetype identifier combining type and vintage (e.g., \texttt{"hospital\_2013"});
\texttt{meta:vintage} records the ASHRAE energy code vintage (\texttt{"pre1980"},
\texttt{"ashrae2004"}, or \texttt{"ashrae2013"}); and \texttt{meta:climateZone} records
an ASHRAE climate zone (2A, 2B, 3C, 4A, 4B, 5A, or 6B) sampled from a floor-area-weighted
distribution derived from the U.S.\ Commercial Buildings Energy Consumption Survey (CBECS).
Each \texttt{HVAC\_Zone} node carries \texttt{meta:floorAreaM2} (zone area in square meters,
drawn from DOE prototype zone area ranges), \texttt{meta:spatialLocation} (one of
\texttt{core}, \texttt{perimeter\_N/S/E/W}), and \texttt{meta:zoneType} (one of 17
functional zone types spanning office, educational, healthcare, and hospitality contexts,
assigned by building type).

\subsubsection{Timeseries References}

Sensor points carry timeseries reference records with unique persistent identifiers
derived from the sensor name and realization index.
Approximately 85\% of sensors in each building carry a timeseries reference, reflecting
realistic deployment scenarios where some sensors may be offline or unregistered.

\section{Methodology and Data Collection}

\begin{figure}[htbp] 
\centering
\includegraphics[width=0.90\linewidth]{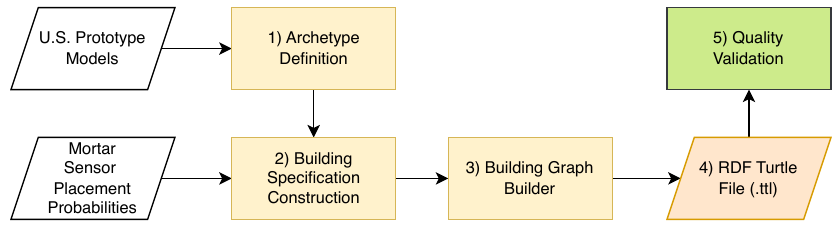}
\caption{BuildGraph generation pipeline. Inputs feed two sequential construction
stages before graph building; the output TTL file undergoes multi-level quality
validation before release.}
\label{fig:pipeline}
\end{figure}

The BuildGraph generation pipeline proceeds through five stages
(Figure~\ref{fig:pipeline}):
(1)~\textit{archetype definition} -- DOE prototype parameters are encoded as Python dataclasses;
(2)~\textit{specification construction} -- each archetype is instantiated into a
\texttt{BuildingSpec} intermediate representation with sampled structural parameters;
(3)~\textit{graph building} -- the Brick builder consumes the spec and produces an
RDF graph with full topology, sensor attachments, and metadata;
(4)~\textit{serialization} -- the graph is written to RDF Turtle; and
(5)~\textit{validation} -- generation-time structural checks and the full SPARQL benchmark
confirm structural completeness (Section~6).
Stages 1--4 and the benchmark design are described in the subsections below.

It is worth stating precisely what BuildGraph inherits and what it contributes. Two elements are
inherited from established sources: the \emph{archetype parameters}---HVAC system type, terminal
type, plant presence, and floor- and zone-count ranges---are taken from the DOE prototype
models, and the \emph{sensor-inclusion priors} are calibrated to the Mortar corpus. Everything
that turns these inputs into queryable knowledge graphs is contributed by this work: the
intermediate building specification and its plausibility constraints, the Brick graph
construction with controlled variation in topology, entity naming, and attachment predicate, the
75-query QAR benchmark, and the released 120-building corpus. Neither DOE nor Mortar supplies
building KGs, benchmark queries, or the controlled structural diversity the dataset provides.

\subsection{Archetype Design}

Each archetype encodes the parameters of one building type and vintage, derived from
the DOE Commercial Prototype Building Models~\citep{doe2024prototype}. The DOE models define
HVAC system type, terminal unit type, chiller and boiler presence, DOAS configuration, and
floor and zone count ranges for each prototype.

The 24 base archetypes (8 building types $\times$ 3 vintages) vary along two axes. First,
\textbf{HVAC system type} determines the central plant configuration and terminal unit type:
VAV and PVAV systems (\textit{medium office}, \textit{large office}, \textit{hospital}) include a
full chilled and hot water plant with VAV terminal boxes; fan coil unit systems
(\textit{hotel}, \textit{outpatient}) use the same plant but with FCU terminals; and
packaged single-zone systems (\textit{small office}, \textit{primary school},
\textit{strip mall}) have no central plant, with each zone served by a dedicated rooftop
unit. Second, \textbf{vintage} encodes energy code requirements that systematically affect
sensor density, metering, and system configuration, as summarized in
Table~\ref{tab:vintages}.

\textbf{Sampling and plausibility bounds.} Within each archetype, structural parameters are
drawn by inclusive uniform sampling from the DOE-derived ranges: floor count, zone count, and
AHU count over their \texttt{(min, max)} bounds, and zone areas by a continuous uniform draw
over per-zone-type ranges (e.g., 30--120~m\textsuperscript{2} for offices, 20--40~m\textsuperscript{2}
for patient rooms). Equipment counts are not drawn blindly: chiller and boiler counts are
conditioned on the archetype's plant-presence flags (and are therefore zero on packaged-unit
buildings), and the sampled AHU count is clamped to at most the zone count, so a building can
never have more air handlers than zones. Because equipment counts are conditioned in this way and
the remaining parameters are drawn within bounded ranges, the sampler cannot construct an
out-of-range or cross-field-inconsistent specification. A schema-validation layer verifies this as
a guard---zone area must be strictly positive, floor count lies in $[1,20]$, and chiller count in
$[0,6]$---together with roughly a dozen cross-field checks (for example, no cooling tower without a
chiller, no chilled-water pumps without a chiller, and no zone above the top floor). These are
construction-time guarantees rather than a post-hoc filter: because the constraints are enforced
during sampling, the guard never rejects a building, and no released building contains an
out-of-range or physically implausible parameter combination.

\subsection{Brick Graph Builder}

\subsubsection{Topology construction}

The builder first creates the spatial hierarchy: floor nodes (\texttt{brick:Floor}) and zone
nodes (\texttt{brick:HVAC\_Zone}) linked via \texttt{brick:hasPart}, with zone type
assignments subject to spatial constraints (lobby zones restricted to the first floor;
mechanical rooms to the first or top floor).

Each AHU feeds its served terminal units via \texttt{brick:feeds} --
VAV boxes with reheat for VAV and PVAV systems, plain \texttt{brick:VAV} terminal
boxes for packaged single-zone (CAV) systems, and FCUs for fan coil systems -- which
in turn deliver conditioned air to their zones (AHU $\rightarrow$ terminal $\rightarrow$ zone).
Brick Schema~1.3 provides no distinct class for PVAV AHUs or CAV terminals; both
are typed using the nearest available class (\texttt{brick:AHU} and \texttt{brick:VAV},
respectively).
Service zones without a terminal unit
(mechanical rooms, server rooms, storage) receive a direct AHU $\rightarrow$ zone
\texttt{brick:feeds} triple to ensure graph reachability. AHU sub-components are
attached via \texttt{brick:hasPart}~\citep{doe2024prototype}; each fan carries a
\texttt{brick:Variable\_Frequency\_Drive} (VFD) via \texttt{brick:controls}. For
central-plant buildings, chilled water and hot water pumps each carry a VFD via
\texttt{brick:controls}; condenser water pumps are constant-speed and do not.

\subsubsection{Sensor and setpoint attachment}

Sensor placement probabilities are derived from empirical analysis of real buildings in the
Mortar corpus~\citep{fierro2019mortar,mortar_resources}. For each equipment class (AHU, VAV, 
FCU, chiller, boiler, cooling tower, supply fan, return fan, cooling coil, heating coil, 
HVAC zone), the fraction of Mortar buildings that carry each sensor subclass was computed to 
establish a base probability. These frequencies were then reviewed against domain knowledge and 
adjusted where the Mortar corpus was thin or unrepresentative -- for example, supplementing 
probabilities for equipment classes with fewer than five Mortar instances. The resulting 
curated probability tables serve as the primary source; an empirical fallback database -- raw 
Brick class frequencies computed directly from the full Mortar corpus without domain review -- covers
sensor subclasses absent from the curated tables.

The vintage density scale (0.40$\times$ for pre-1980, 1.00$\times$ for 2004, 1.40$\times$
for 2013) multiplies each base probability before sampling. For each sensor class, the builder
draws a uniform random variable and includes the sensor if the draw falls below the scaled
probability. Sensors are attached using \texttt{brick:hasPoint}, \texttt{brick:isPointOf},
or both, according to the building's sampled relationship style, reflecting predicate
heterogeneity observed across real Brick deployments~\citep{fierro2019mortar}.
Zone nodes receive direct sensor attachments (zone air temperature, CO$_2$, occupancy
sensors) via the same mechanism, enabling SPARQL queries that traverse from zone to sensor 
without passing through a terminal unit.

\subsection{Reproducibility}

Each building realization is governed by a single integer seed from which two independent
random sequences are derived: a \textit{structural sequence} governing all topology
decisions (zone count, AHU configuration, naming style, relationship style) and a
\textit{sensor sequence} governing probabilistic sensor inclusion draws. Keeping the two
sequences independent ensures structural and sensor variation are separately reproducible --
adjusting sensor probabilities does not alter the topology. Given the same seed, the
generator always produces an identical output file.

\subsection{SPARQL Benchmark Design}
\label{sec:benchmark}

The benchmark comprises 75 SPARQL queries (\texttt{benchmarks/sparql\_queries.json}),
organized into five categories reflecting the types of questions a text-to-SPARQL system
would be expected to answer over building KGs (Table~\ref{tab:benchmark_categories}).

\begin{table}[htbp]
\centering
\caption{SPARQL benchmark query categories}
\label{tab:benchmark_categories}
\begin{tabularx}{\textwidth}{clrX}
\toprule
Cat. & Label & Queries & Example \\
\midrule
1 & Equipment Discovery    & 15 & List all AHUs; find all VAV boxes with reheat \\
2 & Sensor/Point Relations & 15 & Find CO$_2$ sensors on zones; zones with sensors via \texttt{isPointOf} \\
3 & Topological Traversal  & 20 & Chiller $\to$ pump $\to$ AHU chain; floor $\to$ zone $\to$ AHU \\
4 & Aggregation            & 10 & Count terminals per AHU; total zones per floor \\
5 & Multi-hop / Complex    & 15 & 3--5 hop paths, UNION, MINUS operators \\
\bottomrule
\end{tabularx}
\end{table}

Listing~\ref{lst:queries} shows three representative queries spanning the difficulty range,
from a single-class lookup to a five-hop traversal; each carries a natural-language prompt
(shown as a comment) that provides a text-to-SPARQL pair out of the box.

\begin{lstlisting}[caption={Representative benchmark queries; the natural-language prompt appears
as a comment above each query, and ``Category'' denotes the query category (of five).},
label={lst:queries}, basicstyle=\footnotesize\ttfamily]
# Q01 (Category 1): List all air handling units
SELECT DISTINCT ?x WHERE { ?x a brick:AHU }

# Q51 (Category 4): Count terminals served by each AHU
SELECT ?ahu (COUNT(DISTINCT ?t) AS ?n) WHERE {
  ?ahu a brick:AHU ; brick:feeds ?t .
  { ?t a brick:VAV } UNION { ?t a brick:Fan_Coil_Unit }
} GROUP BY ?ahu HAVING (COUNT(DISTINCT ?t) > 0)

# Q75 (Category 5): Chiller-to-zone path reaching
# terminals that carry timeseries sensors
SELECT DISTINCT ?chiller ?zone ?ts WHERE {
  ?chiller a brick:Chiller ; brick:feeds+ ?ahu .
  ?ahu a brick:AHU ; brick:feeds ?terminal .
  ?terminal brick:feeds ?zone .
  ?zone a brick:HVAC_Zone .
  { ?terminal brick:hasPoint ?s }
    UNION { ?s brick:isPointOf ?terminal }
  ?s ref:hasExternalReference ?ref .
  ?ref ref:hasTimeseriesId ?ts
}
\end{lstlisting}

\subsubsection{Scope detection}

Not all queries apply to all buildings: 14 ASK queries auto-detect the features present in
each graph (\texttt{has\_chiller}, \texttt{has\_erv}, \texttt{has\_fcu},
\texttt{has\_zone\_sensors}, \texttt{has\_lobby}, \texttt{has\_mixed\_rel\_style}, and
others), and queries whose required features are absent are marked N/A and excluded from the
QAR denominator. Results are further stratified by four complexity tiers -- T1 (packaged-unit:
small office, primary school, strip mall), T2 (standard central plant: medium office, hotel
pre-1980/2004), T3 (complex VAV+DOAS: large office, hospital/outpatient pre-1980/2004), and
T4 (high-performance DOAS+ERV: hospital, hotel, outpatient 2013) -- to separate
design-driven performance differences from generation failures.

\subsubsection{Query Answerability Rate}

The Query Answerability Rate (QAR) for a set of buildings is defined as:

\begin{equation}
  \text{QAR} = \frac{\sum_{b}\sum_{q} \mathbf{1}[\text{pass}(b,q)]}
               {\sum_{b}\sum_{q} \mathbf{1}[\text{applicable}(b,q)]}
\end{equation}

where $\text{pass}(b,q) = 1$ if query $q$ returns at least one result row for building $b$,
and $\text{applicable}(b,q) = 1$ if all required scope features of $q$ are present in $b$.

An applicable query that returns no rows is scored as a failure: it remains in the denominator
and is excluded from the numerator, so QAR falls. A query marked N/A is removed from both sums
and cannot by itself change QAR. A generator change therefore lowers QAR by turning passing
queries into zero-row failures, not by making queries N/A.

\section{Data Validation and Quality Control}

Dataset quality is validated across six complementary dimensions: generation-time structural
checks and SHACL schema conformance (Section~\ref{sec:generation_validation}), archetype
coverage against national building stock data (Section~\ref{sec:cbecs}), sensor attachment-rate
fidelity against the Mortar corpus (Section~\ref{sec:sensor_coverage}), independent
sensor-composition fidelity against real buildings (Section~\ref{sec:fidelity}), SPARQL benchmark
performance against BuildGraph and real-world Brick corpora
(Sections~\ref{sec:benchmark_results}--\ref{sec:ablation}), and downstream utility via a
text-to-SPARQL cross-building generalization experiment
(Section~\ref{sec:downstream}). Two roles are distinguished among these dimensions: the
generation-time checks are \emph{construction invariants} that guarantee well-formedness by
design and are therefore not independent evidence of quality, whereas the comparisons against
real buildings and national stock data (Sections~\ref{sec:cbecs}, \ref{sec:sensor_coverage},
\ref{sec:fidelity}) and the downstream task (Section~\ref{sec:downstream}) provide
generator-independent evidence.

\subsection{Generation Coverage and Structural Validation}
\label{sec:generation_validation}

The checks in this subsection are \emph{construction invariants}: properties the Brick builder
enforces by design during generation (Section~5), verified here on the serialized output. They
establish that each graph is well-formed and structurally complete, but---because they mirror
the generator's own construction rules---they are a correctness guarantee rather than
independent evidence of dataset quality. The independent, generator-agnostic evidence comes
separately from the sensor-composition comparison against real buildings, the archetype
representativeness check, and the downstream task.

\textbf{Pattern coverage check.} The validator counts how many generated
entity-property-entity triples match entries in the empirical pattern database. A building with
fewer than three matching patterns indicates a structural failure -- a threshold that reflects 
the minimum structural content of a valid building graph: at least one floor-zone, zone-terminal, 
and AHU-terminal relationship must be present. This catches cases such as equipment nodes with no
sensors or zones with no AHU association.

\textbf{Structural completeness checks.} A set of zone-type-aware structural checks and SPARQL
ASK templates is executed against the generated graph immediately after construction:

\begin{enumerate}[leftmargin=0.75in, itemsep=0pt]
  \item Every floor contains at least one zone (\texttt{brick:hasPart})
  \item Every occupied zone is fed by at least one terminal unit (\texttt{brick:feeds});
        service zones (mechanical rooms, storage, server rooms) are excluded from this check
        because they are unconditioned by design
  \item Every AHU feeds at least one downstream node (\texttt{brick:feeds})
  \item Every AHU reaches at least one \texttt{brick:HVAC\_Zone} via \texttt{brick:feeds+}
  \item Chiller $\rightarrow$ Chilled Water Pump $\rightarrow$ AHU chain present when a chiller exists
  \item Every terminal unit feeds at least one zone
  \item Zone sensor attachment via \texttt{brick:hasPoint} (applied to buildings
        with \texttt{hasPoint} or mixed relationship style)
  \item Zone sensor attachment via \texttt{brick:isPointOf} (applied to buildings
        with \texttt{isPointOf} or mixed relationship style)
  \item Every instrumented occupied zone has a \texttt{Zone\_Air\_Temperature\_Sensor};
        zones with occupancy or CO$_2$ sensors but no temperature sensor indicate a
        partial sensor assignment that would be physically inconsistent
\end{enumerate}

Because these invariants are enforced during construction, every generated building satisfies
them by design; the released corpus is produced directly, not selected as the survivors of a
filtering stage. To confirm that the invariants hold in the serialized files, the validator was
run as an independent post-hoc audit over all 120 released buildings: all 120 pass, with
between 4 and 65 matched empirical patterns per building (median 38) and zero structural or
answerability failures. No candidate building was discarded, so no acceptance-rate statistic
applies: the generator emits a balanced corpus of five seeds per archetype, and dataset scale
is bounded only by the number of seeds requested rather than by a pass/fail filter.

Because the released graphs pass these checks by construction, passing does not by itself show that
the checks are discriminating rather than tautological. To demonstrate that they detect genuine
structural deficiency, controlled ablation was applied to the released graphs: each \texttt{brick:feeds},
\texttt{brick:hasPart}, \texttt{brick:hasPoint}, and \texttt{brick:isPointOf} triple was removed
independently with probability $p$, and the ablated graphs were re-validated (one seed per
archetype; the deposited dataset is unchanged). At $p{=}0$ all graphs pass, reproducing the
construction guarantee; removing only $5\%$ of these edges ($p{=}0.05$) already causes $62.5\%$
of graphs to be rejected, rising to $100\%$ at $p{=}0.5$ (Table~\ref{tab:reject}). Violations of
the structural invariants dominate, with query-answerability failures growing as more of the
graph is removed. The checks therefore detect precisely the missing-structure defects
characteristic of incomplete models, confirming that passing them is an informative property
rather than a tautology.

\begin{table}[htbp]
\centering
\caption{Validator reject rate under controlled structural ablation of the released graphs
         (24 buildings, one seed per archetype). Each \texttt{feeds}/\texttt{hasPart}/\texttt{hasPoint}/\texttt{isPointOf} triple is removed independently with probability
         $p$; the deposited dataset is unchanged.}
\label{tab:reject}
\begin{tabular}{ccc}
\toprule
Ablation $p$ & Reject rate & Dominant failure \\
\midrule
0.00 & \phantom{0}0.0\% & --- (all pass) \\
0.05 & 62.5\% & structural \\
0.10 & 83.3\% & structural \\
0.20 & 91.7\% & structural \\
0.30 & 95.8\% & structural \\
0.50 & 100.0\% & structural + answerability \\
\bottomrule
\end{tabular}
\end{table}

\subsubsection{SHACL Conformance}
\label{sec:shacl}

Beyond structural checks, schema correctness is verified using the SHACL shapes bundled with
Brick Schema 1.3~\citep{fierro2022brick}, which introduced formal SHACL constraint shapes
alongside wider class coverage and alignment mappings to related ontologies. Running the Brick SHACL shapes
via \texttt{pyshacl}~\citep{pyshacl} against the full set of 120 released buildings -- drawn
from all archetype and vintage combinations -- yields a 100\% conformance rate
(0 shape violations across all files). Reproducibility was confirmed by generating the
released dataset twice using the same seed sequence and verifying identical triple sets on
every output file.

\subsection{Archetype Representativeness}
\label{sec:cbecs}

The 24 archetypes are validated against real-world commercial building stock using the
U.S.\ Energy Information Administration (EIA) Commercial Buildings Energy Consumption Survey (CBECS) 
2018 Public Use Microdata,
a nationally representative survey of 6{,}436 buildings representing 5.92 million commercial
buildings in the United States~\citep{eia_cbecs_2018}.

The BuildGraph archetypes collectively account for \textbf{30.2\%} of the national
commercial building stock by count (Table~\ref{tab:cbecs_coverage}). CBECS confirms
HVAC system class alignment for offices, K-12, and retail, where packaged DX/RTU is the
dominant system, and for hospitals, where chilled water leads. For hotel and outpatient
archetypes, CBECS shows packaged DX/RTU as the plurality; BuildGraph follows the DOE
prototype FCU-with-central-plant assignment, which targets mid-to-large facilities where
central chilled water is standard. The three vintages together span the full U.S.\
commercial building age distribution. CBECS validates archetype selection and HVAC class
alignment; sensor placement and graph topology are addressed in subsequent subsections.

\begin{table}[htbp]
\centering
\caption{CBECS 2018 coverage by BuildGraph archetype group. Vintage stock is the
weighted share of buildings by year-of-construction category
(pre-1980 / 1980--2012 / 2013--2018).}
\label{tab:cbecs_coverage}
\begin{tabularx}{\textwidth}{Xrrr}
\toprule
Archetype group & CBECS share & Primary cooling & Vintage stock \\
                &             &                 & (pre-80 / 80--12 / 13+) \\
\midrule
Office (small/medium/large)  & 15.6\% & Packaged DX/RTU (82.8\%) & 47\% / 48\% /  5\% \\
Education (K-12)    &  6.4\% & Packaged DX/RTU (72.7\%) & 50\% / 46\% /  4\% \\
Lodging (hotel)     &  2.3\% & Packaged DX/RTU (84.0\%) & 33\% / 56\% / 11\% \\
Retail (strip mall) &  2.8\% & Packaged DX/RTU (89.3\%) & 45\% / 47\% /  8\% \\
Health (outpatient) &  3.0\% & Packaged DX/RTU (80.8\%) & 32\% / 63\% /  5\% \\
Health (hospital)   &  0.1\% & Chilled water (55.2\%)   & 41\% / 51\% /  8\% \\
\midrule
\textbf{Total}      & \textbf{30.2\%} & & \\
\bottomrule
\end{tabularx}
\end{table}

\textbf{Included and excluded categories.} The eight archetypes are the DOE Commercial Prototype
types that are both empirically calibrated to U.S.\ stock and dominant by building count,
together representing 30.2\% of commercial buildings. The principal CBECS categories not
modeled---warehouse and storage, food service and food sales, public assembly, religious worship,
and miscellaneous service---differ chiefly in envelope and internal loads but reuse the same HVAC
families already covered here (packaged DX/RTU, VAV with a central plant, and fan-coil systems);
a warehouse or big-box retail space, for example, is served by packaged units structurally
analogous to the strip-mall and small-office archetypes. Because BuildGraph's contribution is the
KG \emph{structure}---equipment topology, sensor attachment, and query patterns---rather than
category-specific thermal behavior, the structural patterns represented here carry over to the
excluded categories, so the dataset's representational coverage of KG structure extends well beyond
its share of buildings by count and matches the stated uses (text-to-SPARQL, ontology benchmarking,
and portable-analytics testing). Coverage is moreover not fixed at the deposited corpus: the open
generator allows any excluded category to be produced on demand by supplying its DOE-style
parameters, so the released set is a starting point the community can extend rather than a ceiling.
The exclusions are noted as a coverage limitation for applications that depend on category-specific
instrumentation.

\subsection{Sensor Attachment Rate Validation}
\label{sec:sensor_coverage}

The Mortar empirical probabilities used as input to the generator (Section~5)
are derived from the fraction of real buildings that carry each sensor type on each
equipment class.
To verify that the generator faithfully realizes these probabilities in the output distribution,
each generated building is queried for the fraction of equipment instances that carry each
key sensor and compare the measured corpus-wide rate to the input probability used by
the generator.

Table~\ref{tab:sensor_coverage} reports the comparison for the 40 standard-density buildings
(2004 vintage, whose density scale factor is 1.00$\times$, enabling direct comparison between
input probabilities and output rates without a scaling correction;
190 AHU instances, 443 VAV instances, 1{,}014 zone instances).
Output rates are within $\pm 3$~percentage points of input probabilities for all AHU and VAV
trunk sensors. The Chiller shows a larger deviation ($-9$~pp), reflecting that only five of eight building
types include a central chiller and the small sample (36 instances) widens the confidence
interval.
Zone-level IAQ sensors (CO\textsubscript{2}, occupancy) match their input probabilities
within $\pm 3$~pp.

\begin{table}[htbp]
\centering
\caption{Sensor attachment rate: Mortar input probability vs.\ BuildGraph output rate
  (40 standard-density buildings, 2004 vintage).
  Output rate = fraction of equipment instances carrying the point type
  via \texttt{brick:hasPoint} or \texttt{brick:isPointOf}.
}
\label{tab:sensor_coverage}
\small
\begin{tabular}{llrrl}
\toprule
Equipment & Point Type & Input & Output & $\Delta$ \\
\midrule
AHU & Supply\_Air\_Temperature\_Sensor        & 92\% & 95\% & $+3$~pp \\
AHU & Return\_Air\_Temperature\_Sensor        & 82\% & 82\% & $\pm 0$~pp \\
AHU & Supply\_Air\_Flow\_Sensor               & 78\% & 77\% & $-1$~pp \\
AHU & Supply\_Air\_Static\_Pressure\_Sensor   & 82\% & 81\% & $-1$~pp \\
VAV & Zone\_Air\_Temperature\_Sensor          & 92\% & 94\% & $+2$~pp \\
VAV & Supply\_Air\_Flow\_Sensor               & 82\% & 80\% & $-2$~pp \\
VAV & Zone\_Air\_Cooling\_Temperature\_Setpoint & 88\% & 89\% & $+1$~pp \\
Chiller & Chilled\_Water\_Supply\_Temperature\_Sensor & 95\% & 86\% & $-9$~pp \\
HVAC\_Zone & CO\textsubscript{2}\_Sensor      & 38\% & 41\% & $+3$~pp \\
HVAC\_Zone & Occupancy\_Sensor                & 42\% & 40\% & $-2$~pp \\
\bottomrule
\end{tabular}
\end{table}

\subsection{Independent Fidelity: Sensor Composition vs.\ Real Buildings}
\label{sec:fidelity}

The validation described so far establishes that the generated graphs are
\emph{structurally complete} and that the generator faithfully realizes its input
sensor-inclusion probabilities. Neither, however, is independent of the generator:
the structural checks (Section~\ref{sec:generation_validation}) verify the
construction invariants the builder enforces by design, and the attachment-rate
comparison (Section~\ref{sec:sensor_coverage}) compares the output against the same
probabilities used to produce it. To provide evidence that does not depend on the
generator's own rules, BuildGraph is compared against a corpus of real buildings on a
property the generator does not directly target: the \emph{composition} of point types
attached to each equipment class---the relative proportion of each point type among all
points on instances of that class. The reference is the set of real Brick buildings
published with the Mortar paper and made publicly available by the Brick Schema
project~\citep{fierro2019mortar,mortar_resources}, queried here at the level of realized
per-equipment point mixtures rather than inclusion frequencies.

Real building models are structurally sparser than BuildGraph: a real air-handling unit
in the corpus carries a median of six points, whereas a BuildGraph AHU carries fourteen.
A direct comparison of point counts therefore reflects the well-documented incompleteness
of deployed models rather than the plausibility of the synthetic ones. Instead, the point-type
\emph{proportions} are compared, which are invariant to how densely a given unit is
instrumented, and two questions are separated. First, restricted to the point types that both
corpora model, do the relative proportions agree (a test of realism)? Second, how much of
each corpus's instrumentation falls outside that shared vocabulary (a measure of the
completeness gap)?

Table~\ref{tab:fidelity} reports both. For the point-type vocabulary common to both corpora,
the compositional proportions agree closely---cosine similarity of $0.94$, $0.91$, and
$0.86$ for air-handling units, variable-air-volume terminals, and chillers, respectively.
The agreement holds for the core physical measurements: AHU return-air, mixed-air, and
supply-air static-pressure sensors each account for roughly $5$--$6\%$ of AHU points in both
corpora, and chilled-water supply and return temperature sensors each account for roughly
$10$--$11\%$ of chiller points in both. Where a point type appears in real buildings, it
appears in BuildGraph at a comparable relative frequency.

The two corpora diverge along the shared-coverage dimension, and the divergence is by design.
The common vocabulary accounts for $40$--$48\%$ of BuildGraph's AHU and VAV points; the
remainder are setpoints and secondary commands (e.g., zone heating and cooling setpoints,
damper-position commands) that BuildGraph models but that heterogeneous real deployments
frequently omit. Conversely, the real corpus assigns a substantial share of points to coarse,
legacy classes (e.g., a generic \texttt{brick:Command}) that BuildGraph resolves to specific
subclasses. Both differences reflect BuildGraph's greater completeness and vocabulary
specificity---the properties the dataset is designed to provide---rather than a departure
from realistic sensor placement. Zone-level points are excluded from this analysis: real
zones in the corpus carry essentially only a temperature sensor, so the shared vocabulary
reduces to a single type and the comparison is uninformative; the richer zone instrumentation
BuildGraph provides (heating and cooling setpoints, occupancy, and CO\textsubscript{2}
sensors) is precisely the completeness that real zone models lack.

In summary, this analysis supports two claims that QAR alone cannot: conditional on the
sensor vocabulary that real buildings model, BuildGraph reproduces real-world sensor-mixture
proportions; and BuildGraph is systematically more complete than publicly available real
models. It does not establish that individual synthetic buildings are indistinguishable from
specific real facilities, which remains outside the scope of a distributional grounding.

\begin{table}[htbp]
\centering
\caption{Sensor-composition fidelity against real Mortar buildings. \emph{Composition
         similarity} is the cosine similarity of point-type proportions, restricted to point
         types both corpora model (density- and vocabulary-normalized); \emph{shared coverage}
         is the share of each corpus's points that fall within that common vocabulary.}
\label{tab:fidelity}
\begin{tabular}{lrrrcc}
\toprule
Equipment & \multicolumn{2}{c}{Instances} & Shared & Composition & Shared coverage \\
 & Real & BuildGraph & types & similarity & (BG / Real) \\
\midrule
AHU & 185 & 545 & 6 & 0.94 & 40\% / 54\% \\
VAV & 1992 & 1280 & 8 & 0.91 & 48\% / 67\% \\
Chiller & 38 & 102 & 10 & 0.86 & 90\% / 51\% \\
\bottomrule
\end{tabular}
\end{table}

\subsection{Benchmark Results}
\label{sec:benchmark_results}

Table~\ref{tab:qar_main} reports QAR by tier and query category for the full released dataset
(120 buildings; 5 seeds $\times$ 24 archetypes).

\begin{table}[htbp]
\centering
\caption{QAR by tier and query category (120 released buildings,
         5 seeds $\times$ 24 archetypes)}
\label{tab:qar_main}
\begin{tabular}{lccccc}
\toprule
Category & T1 & T2 & T3 & T4 & Overall \\
\midrule
1. Equipment Discovery    & 95.5\% & 96.8\% & 96.6\% & 100.0\% & 96.9\% \\
2. Sensor/Point Relations & 77.1\% & 83.2\% & 84.6\% &  90.2\% & 82.6\% \\
3. Topol.\ Traversal      & 89.0\% & 91.8\% & 93.4\% & 100.0\% & 92.5\% \\
4. Aggregation            & 97.3\% & 98.8\% & 99.7\% & 100.0\% & 98.7\% \\
5. Multi-hop / Complex    & 82.6\% & 90.3\% & 92.6\% &  92.3\% & 89.1\% \\
\midrule
\textbf{Overall QAR} & \textbf{87.5\%} & \textbf{91.7\%} & \textbf{92.9\%} &
                       \textbf{96.5\%} & \textbf{91.5\%} \\
\bottomrule
\end{tabular}
\end{table}

Category~2 (Sensor/Point Relations) scores lower than other categories because three queries
(Q17, Q18, Q29) use \texttt{brick:isPointOf} as the primary attachment direction, firing only
on buildings with an inverse or mixed relationship style; a fourth query (Q30) is scoped to
the mixed style only. Pass rates are consistent with the fraction of buildings using each
style by design, and reflect an intentional feature: a trained model must handle both
\texttt{hasPoint} and \texttt{isPointOf} patterns as observed in real Brick deployments.

\subsection{Comparison with Real-World Brick Buildings}

To establish whether the benchmark queries are answerable in practice and to contextualize the
synthetic QAR figures, the same 75-query benchmark was run against a corpus of 59 publicly
available Brick building files drawn from two sources: (1)~the 45 anonymized Brick models
published with the Mortar paper and made publicly available by the Brick Schema
project~\citep{fierro2019mortar,mortar_resources} (\texttt{bldg1}--\texttt{bldg44} and
\texttt{smc}); and (2)~14 additional community and reference Brick
models~\citep{brickschema_examples}, including the Soda~Hall, EBU3B, and Rice reference models and
an ASHRAE Guideline~36~\citep{ashrae_g36} model. All are drawn from public Brick model
collections, so the comparison can be reproduced from openly available files. All figures use aggregate QAR
(total passing queries / total applicable queries),
consistent with the BuildGraph metric. Scope detection was applied identically to all real buildings;
results are reported in Table~\ref{tab:qar_real}.

\begin{table}[htbp]
\centering
\caption{QAR comparison: 59-building Brick corpus and BuildGraph.
         All figures are aggregate QAR (total passing / total applicable queries).}
\label{tab:qar_real}
\begin{tabular}{lrr}
\toprule
Category & Brick corpus & BuildGraph \\
         & (agg., $n{=}59$) & ($n{=}120$) \\
\midrule
1.\ Equipment Discovery    & 42.3\% & 96.9\% \\
2.\ Sensor/Point Relations & 21.2\% & 82.6\% \\
3.\ Topol.\ Traversal      & 41.6\% & 92.5\% \\
4.\ Aggregation            & 56.1\% & 98.7\% \\
5.\ Multi-hop / Complex    & 15.3\% & 89.1\% \\
\midrule
\textbf{Overall QAR}      & \textbf{35.8\%} & \textbf{91.5\%} \\
\bottomrule
\end{tabular}
\end{table}

\noindent Across the 59-building corpus, aggregate QAR is 35.8\%, a 55.7-point gap below
BuildGraph (91.5\%). The gap measures \emph{structural completeness}: the generated graphs contain
the query-relevant relationships the benchmark exercises, whereas real deployments, assembled
incrementally and rarely as end-to-end facility models, often do not. This completeness is what
makes BuildGraph valuable to the community as a development and benchmarking substrate. A
text-to-SPARQL model, a portable-analytics application, or an ontology tool can be exercised against
graphs that are guaranteed to contain every relationship a query requires, so that a failure can be
attributed to the method rather than to missing data---a controlled reference that incomplete real
corpora cannot provide, and against which the completeness of any given real building can itself be
measured. The comparison is therefore a statement about completeness, not realism; sensor-placement
realism is assessed separately in Section~\ref{sec:fidelity}.
This incremental assembly follows a recognizable pattern. Real Brick models are built from
heterogeneous data sources -- building automation system point lists, equipment schedules, and
schematic diagrams -- using annotation tools that each address a specific system layer. Air-side equipment (AHUs, VAV terminals, and
associated sensors) is typically modeled first because it is most directly relevant to
operational and control applications. Central plant components (chillers, boilers, and
associated pumping systems), distribution topology, and sensor-to-equipment attachment links
are frequently absent, either because the annotation pipeline did not extend to those layers
or because the original modeling objective did not require them. A building can be physically
complex and fully instrumented yet score poorly on the benchmark if only a subset of its
systems has been semantically modeled in Brick form.

Category~4 (Aggregation) scores highest at 56.1\%, while Category~5 (Multi-hop / Complex)
is the weakest at 15.3\%. Five files score 0\% (no HVAC equipment in Brick form); the
highest-scoring file (62.2\%) is an ASHRAE Guideline~36~\citep{ashrae_g36} reference model
designed for structural completeness, not a real deployed building.

The category with the largest discrepancy is Multi-hop / Complex (Category~5): 15.3\% for
the corpus vs.\ 89.1\% for BuildGraph, a 73.8-point gap. End-to-end chiller-to-zone
traversals require every equipment layer to be present simultaneously; current annotation
workflows do not achieve this because plant, distribution, and terminal systems are typically
modeled by separate tools or omitted entirely.
BuildGraph is designed to fill this gap: a structurally complete dataset enables text-to-SPARQL
systems to be evaluated and trained on query patterns that real KGs cannot currently exercise.

\subsection{Component Contribution Analysis}
\label{sec:ablation}

To quantify the contribution of each BuildGraph design component to dataset quality, QAR is
measured at four pipeline stages across the same 120 buildings (Table~\ref{tab:ablation}).

\begin{table}[htbp]
\centering
\caption{QAR ablation by pipeline stage. Each row adds one sensor-related design component to
         the previous stage. Relationship style is held at the released setting in all
         stages, so the comparison isolates sensor density and placement; Stage~4 is the
         full released dataset.}
\label{tab:ablation}
\small\setlength{\tabcolsep}{4pt}
\begin{tabular}{lcccccc}
  \toprule
  Pipeline stage & Category 1 & Category 2 & Category 3 & Category 4 & Category 5 & Overall \\
  \midrule
  1.\ DOE structure only (no sensors) & 96.9\% &  0.0\% & 54.8\% & 61.4\% & 26.8\% & 51.8\% \\
  2.\ +Uniform sensors ($p=0.5$)      & 96.9\% & 82.2\% & 94.1\% & 99.4\% & 90.1\% & 91.9\% \\
  3.\ +Mortar-calibrated sensors      & 96.9\% & 87.7\% & 97.7\% & 99.9\% & 90.7\% & 94.3\% \\
  4.\ +Vintage scaling \& rel\_style  & 96.9\% & 82.6\% & 92.5\% & 98.7\% & 89.1\% & 91.5\% \\
  \bottomrule
\end{tabular}
\end{table}

DOE-grounded topology alone (Stage~1) yields near-perfect equipment discovery (Category~1:
96.9\%) but fails on sensor-dependent queries: Category~2 is 0\% and overall QAR is only
51.8\%, confirming that topology is a necessary scaffold but sensor attachment drives the
majority of quality gains.

Adding sensors at a flat 50\% probability (Stage~2) produces the largest single gain
(+40.1~pp to 91.9\%). Categories~3 and~5 also rise sharply because many traversal and
multi-hop queries require a sensor anchor along the path; correct topology alone leaves
them unanswerable.

Switching to Mortar-calibrated probabilities (Stage~3) adds 2.4~pp overall (91.9\%
$\to$ 94.3\%). The modest increment reflects the binary nature of QAR; Mortar calibration
matters more for distributional realism -- matching observed sensor density -- than for
aggregate answerability.

Stage~4 (full BuildGraph) is the only stage that scores \textit{below} Stage~3
(94.3\% $\to$ 91.5\%, $-2.8$~pp). This is a deliberate design choice, not a regression.
Stage~3 applies one sensor density to every building; Stage~4 adds vintage density
scaling, so pre-1980 buildings are sparsely instrumented and 2013 buildings densely
instrumented. The sparser buildings leave more applicable queries with no matching rows, and
those queries are scored as failures, which lowers aggregate QAR while making the dataset more
representative of real building stock. A text-to-SPARQL model trained only on Stage~3 would
overfit to uniformly dense graphs and fail on the sparse buildings common in real deployments.

The reduction is entirely a shift from passing to failing queries. Between Stages~3 and~4 the
Category~2 N/A count is unchanged at 273 query--building pairs and the applicable denominator is
identical at 1{,}527; the 5.1-point decrease consists of 77 pairs that move from pass to fail.

Measured by vintage, the decrease is confined to the sparse vintage. Aggregate QAR is 83.4\% for
pre-1980 against 95.1\% and 95.8\% for 2004 and 2013---an 11.7-point spread concentrated in
Category~2 (70.9\% pre-1980 vs.\ 88--89\% later) and Category~3 (79.9\% vs.\ 98--99\%), where
fewer attached sensors leave applicable queries with no matching rows. Category~2 falls from
86.6\% to 70.9\% in the pre-1980 vintage between Stages~3 and~4 and is unchanged in the 2004 and
2013 vintages (88.0\% and 88.4--89.0\%), confirming density scaling as the sole driver.

Predicate style is a separate effect on the level of Category~2 in every stage. Three benchmark
queries are written with a single attachment predicate---Q17 with \texttt{brick:hasPoint}, Q18
and Q29 with \texttt{brick:isPointOf}---and return no rows on graphs built with the opposite
style. These queries remain in scope and are scored as failures; their combined QAR is 54.2\% in
both the 2004 and 2013 vintages at Stages~3 and~4 alike, and excluding them Category~2 reaches
98.5\% and 99.7\%. Predicate diversity therefore sets a constant ceiling on Category~2 across
stages and does not contribute to the Stage-4 decrement.

\subsection{Downstream Validation: Text-to-SPARQL Generalization}
\label{sec:downstream}

A key motivation for BuildGraph is enabling systems that translate natural language building
questions into SPARQL queries, allowing building operators, energy managers, and researchers
to retrieve information from a building knowledge graph without writing SPARQL by hand. For
such a system to generalize across buildings -- producing correct queries whether the target
is a hospital or a small office, regardless of how its KG is structured or named -- it
requires diverse, structurally complete building data, whether used as few-shot examples
or an evaluation benchmark. To test both of these uses, Gemma~4 was evaluated (the instruction-tuned
\texttt{gemma4:26b} checkpoint---a 26B-parameter Mixture-of-Experts model activating roughly 4B
parameters per token, released by Google DeepMind and served locally through the Ollama
runtime)~\citep{gemma4},
a general-purpose large language model not
specialized for buildings or SPARQL, on 12 held-out buildings (3 from each complexity tier,
T1 through T4, where T1 corresponds to simpler packaged-unit buildings and T4 to complex DOAS
systems with energy recovery). The remaining 108 buildings served as the pool for
cross-building examples.

Each building is presented to the model as a structured natural language summary extracted
from its RDF graph, describing the equipment inventory (e.g., ``3 AHUs, 18 VAV terminals,
2 chillers''), plant-to-zone service topology, sensor types per equipment class, and entity
naming convention -- providing sufficient structural information to construct a valid SPARQL
query without accessing the raw RDF file. A structured summary is supplied rather than the raw
Turtle because a dense building graph (up to roughly 12{,}800 triples) exceeds a practical
prompt budget; this is a deliberate simplification of the task, so the results should be read
as a conservative lower bound on what the same model could achieve with direct or
retrieval-augmented access to the graph. This limitation is revisited in
Section~\ref{sec:limitations}.

\textbf{Prompting conditions.} Two conditions were evaluated:

\textit{Zero-shot}: The model receives only the building summary and the question, with no
worked examples. It relies entirely on its pre-trained knowledge of SPARQL syntax and the
Brick Schema vocabulary.

\textit{3-shot}: The model additionally receives three complete worked examples -- each
pairing a building summary and a natural language question with the correct reference SPARQL
query -- drawn from \textit{other} buildings of comparable complexity. These cross-building
examples show the model how Brick class names, entity identifiers, and query structures
appear in practice.

\textbf{Evaluation metric.} SPARQL queries return tabular results -- sets of rows where each
row represents a matching entity or value (e.g., a list of AHU identifiers, or a count of
sensors per zone). Row-Matching~F1~\citep{buildingqa} measures correctness by comparing the
rows returned by the model's query against those returned by the reference query: 100\% means
exactly the right results were retrieved; lower scores reflect missing or spurious rows.
Scores are computed only over queries for which the reference query returns at least one
result in the test building -- i.e., queries the building structurally contains the answer
to. The QAR ceiling column in Table~\ref{tab:nl2sparql} reports this fraction; a ceiling of
97.1\% confirms that F1 scores reflect model capability rather than data incompleteness.

\textbf{Results.} Table~\ref{tab:nl2sparql} reports Row-Matching~F1 by query category. The
model always produces syntactically valid SPARQL; all errors are in query logic, confirming
that difficulty lies in mapping natural language intent onto Brick Schema structure rather
than in generating parseable SPARQL. Overall F1 improves from 9.8\% zero-shot to 27.3\%
with worked examples (+17.5~pp), demonstrating that BuildGraph examples are genuinely useful
as cross-building context.

\begin{table}[htbp]
  \centering
  \caption{Text-to-SPARQL Row-Matching F1 by query category. Gemma~4 (26B),
           zero-shot vs.\ 3-shot cross-building prompting on 12 held-out buildings.
           QAR ceiling is the aggregate QAR of the test set.}
  \label{tab:nl2sparql}
  \begin{tabular}{lccc}
    \toprule
    Category & Zero-shot & 3-shot & QAR ceiling \\
    \midrule
    1.\ Equipment Discovery    & 29.0\% & 63.0\% & 100.0\% \\
    2.\ Sensor/Point Relations &  7.1\% &  4.1\% &  95.1\% \\
    3.\ Topol.\ Traversal      &  4.7\% & 20.3\% &  99.0\% \\
    4.\ Aggregation            & 10.8\% & 48.6\% & 100.0\% \\
    5.\ Multi-hop / Complex    &  0.0\% & 10.1\% &  91.8\% \\
    \midrule
    \textbf{Overall}           & \textbf{9.8\%} & \textbf{27.3\%} & \textbf{97.1\%} \\
    \bottomrule
  \end{tabular}
\end{table}

The results establish two findings relevant to BuildGraph's utility. First, the dataset
provides genuinely challenging test cases: Category~5 (Multi-hop / Complex) scores 0.0\%
zero-shot and 10.1\% with examples, confirming that queries requiring end-to-end traversal
across multiple connected equipment layers remain difficult even with cross-building context.
The persistent gap between the 3-shot score (27.3\%) and the QAR ceiling (97.1\%) reflects
model limitations -- nearly all queries are structurally answerable, but the model does not
yet reliably generate the correct traversal patterns. Second, cross-building examples are
most effective for aggregation queries: the +37.8~pp gain on Category~4
(10.8\%~$\to$~48.6\%) shows that examples teach counting and grouping patterns the model
cannot infer from a building description alone. T1 buildings (packaged-unit systems) reach
31.3\% F1 with examples while T4 buildings (complex DOAS systems) reach 24.4\%, reflecting
the greater query difficulty at higher complexity tiers.

Category~2 (Sensor/Point Relations) is the hardest query class, and unlike the others it does not
benefit from the few-shot examples (mean Row-Matching F1 of 7.1\% zero-shot and 4.1\% with three
examples). This follows from the structure of the task rather than being anomalous. A correct
sensor-relationship query must jointly resolve three requirements that the other categories do not
impose together: the specific sensor class; the direction of the sensor-to-equipment
attachment---the benchmark deliberately includes both export conventions, with
\texttt{brick:hasPoint}-style and \texttt{brick:isPointOf}-style variants of the same question (for
example, Q17 and Q18)---and the result projection. The few-shot exemplars are, moreover, selected
from Categories~1, 3, and 5---equipment discovery, topology traversal, and complex cross-domain
queries---and never from Category~2 itself, so the model is never shown a sensor-relationship example
before answering one. This account is borne out by a fully deterministic (temperature~0) replication
of the Category~2 evaluation, in which the few-shot condition improved none of the 155 held-out
(building, query) pairs over the zero-shot condition and degraded sixteen, confirming that the
regression is systematic rather than an artifact of stochastic decoding. Presented with exemplars
from other classes, the model adopts projection and attachment-direction conventions that do not
match those the sensor-relationship questions require; consequently the generated Category~2 queries
fail as \emph{executable-but-wrong}---valid SPARQL that misresolves the projection or attachment
direction---rather than as malformed queries. Predicate- and projection-aware example selection is
therefore the natural direction for improving this class.

The twelve held-out buildings are stratified across all four complexity tiers (three per
tier), spanning the full difficulty range from packaged-unit to DOAS systems, and each is
evaluated on the complete 75-query benchmark; the remaining 108 buildings form the
cross-building example pool. This design directly tests the property the dataset is built to
provide---generalization to unseen buildings---through the two uses demonstrated here,
few-shot examples and an evaluation benchmark. Broader evaluations across additional models
and downstream applications are natural complements to this result.

\section{Usage Notes}

BuildGraph's primary contribution is a reusable artifact---an openly released corpus, generator,
and benchmark---designed to be used and extended by others rather than to establish a new method or
a single empirical finding. The following notes describe the use cases the dataset supports, how to
extend it, and its limitations.

\subsection{Intended Use Cases}

\textbf{Text-to-SPARQL development and evaluation.} The dataset supports systems that
translate natural language building queries into SPARQL, serving as few-shot examples for
in-context prompting, training data for model fine-tuning, and an evaluation benchmark for
cross-building generalization. Each
of the 75 benchmark queries already carries a natural language description, providing 75
text-to-SPARQL pairs out of the box. The practical bottleneck for training is \textit{paraphrase
diversity} -- multiple phrasings per query template -- which can be addressed by drawing on
BuildingQA's question vocabulary~\citep{buildingqa} or automated generation. The diversity in URI naming conventions,
relationship styles, and building topologies across 120 buildings ensures that a model trained
on BuildGraph does not overfit to a single structural pattern.

\textbf{Building ontology benchmarking.} The dataset can serve as a reference corpus for
evaluating new building ontology tools, reasoners, or schema validators. The QAR benchmark
provides a reusable evaluation protocol applicable to any Brick-formatted building KG.

\textbf{Cross-building analytics testing.} The 120-building corpus provides a ready-made
testbed for evaluating portable SPARQL analytics -- fault detection rules, commissioning
scripts, and energy benchmarking queries -- across diverse building types, HVAC
configurations, and sensor densities. Because all buildings share a common Brick vocabulary,
a query written for one building can be executed across the full corpus without modification,
enabling systematic evaluation of how well an analytics application generalizes across
building types and vintages.

\textbf{Building domain model development.} The structured metadata is also useful for any
model that reasons about HVAC systems, equipment relationships, or sensor coverage -- beyond
text-to-SPARQL. The scale (120 buildings, 83 Brick classes, controlled variation in topology
and instrumentation) makes BuildGraph suitable as domain-specific training data, few-shot examples, or evaluation
benchmarks for building-domain language and reasoning models.

\subsection{Extending the Dataset}

The released 120 buildings represent five seeds per archetype. Researchers needing greater
scale or diversity can generate additional realizations using the open-source BuildGraph
generator: any integer seed produces a new, reproducible building
instance with independently sampled topology, naming style, and sensor configuration. New
building archetypes can be introduced by defining an archetype dataclass with DOE-style
parameters (floor count ranges, HVAC system type, zone type distribution, and terminal unit
configuration) without modifying the Brick builder or benchmark infrastructure. The sensor
probability tables can be recalibrated for specific deployment contexts -- for example, to
reflect denser instrumentation in research buildings or sparser coverage in older facilities
-- by adjusting the empirical pattern database prior to generation.

\subsection{Known Limitations}
\label{sec:limitations}

\textbf{Downstream evaluation uses graph summaries.} The text-to-SPARQL experiment
(Section~\ref{sec:downstream}) presents each building to the model as a structured
natural-language summary rather than the raw RDF graph, a simplification adopted to keep dense
graphs within the model's context window. The reported scores should therefore be read as a
conservative lower bound; evaluating models with direct or retrieval-augmented access to the
full graph, and over the full 120-building corpus, is left to future work.

\textbf{No sensor timeseries data.} The dataset represents structural and topological building
information only. Sensor nodes carry timeseries reference identifiers but no actual measurement
values. Applications requiring time-series data must supplement the KG with a separate data
source.

\textbf{Scope limited to topology.} BACnet references, DDC controller assignments, schedule
bindings, equipment sizing (chiller tonnage, design airflow), and efficiency ratings (COP, EER)
are out of scope; the dataset models the structural and topological layer only.

\textbf{Brick Schema 1.3 only.} All files use the Brick~1.3 class hierarchy and namespace
(\url{https://brickschema.org/schema/Brick\#}). Tools or validators built against earlier
versions of the schema (e.g., Brick~1.1) may report unknown classes for equipment types
introduced in later releases. Brick~1.4 is also available and introduces additional class
refinements and updated SHACL shapes; the namespace is unchanged, so TTL files remain
forward-compatible, but users should re-run SHACL validation against the Brick~1.4 shape
library if applying updated constraint rules.

\textbf{Synthetic topology.} While grounded in empirical patterns and DOE prototypes, the
generated buildings do not represent any specific real facility. Statistical properties match
real buildings at the distribution level; individual graphs should not be used as substitutes
for site-specific building models.

\availability{Some or all data, models, or code generated or used during the study are available in a
repository online. The dataset is deposited at Zenodo
(\url{https://doi.org/10.5281/zenodo.20015123}), and the generator source code, benchmark
queries, and pattern databases are available in the BuildGraph repository; both are released
under a CC~BY-NC~4.0 license. All data used to produce the dataset (DOE prototype parameters,
empirical pattern databases) are documented in the repository.}
\acknowledgments{No funding was received in support of this study.}

\end{document}